# On the Impact of ISA Extension on Energy Consumption of I-Cache in Extensible Processors

Noushin Behboudi, Mehdi Kamal, Ali Afzali-Kusha
School of Electrical and Computer Engineering, University of Tehran
{nshn.behboudi, mehdikamal, afzali}@ut.ac.ir

*Abstract*— **As is widely known, the computational speed and power consumption are two critical parameters in microprocessor design. A solution for these issues is the application specific instruction set processor (ASIP) methodology, which can improve speed and reduce power consumption of the general purpose processor (GPP) technique. In ASIP, changing the instruction set architecture (ISA) of the processor will lead to alter the number and the mean time of accesses to the cache memory. This issue has a direct impact on the processor energy consumption. In this work, we study the impacts of extended ISA on the energy consumption of the extended ISA processor. Also, we demonstrate the extended ISA let the designer to reduce the cache size in order to minimize the energy consumption while meeting performance constraint.**

## I. INTRODUCTION

Power consumption has become a hot topic in digital design over the past decades due to its important role in portable devices. Additionally, as a result of increasing the demands for the high performance systems, the power consumption of the digital systems is increased which leads to higher temperature. Increasing the temperature has an obvious deteriorating effect on performance and also the reliability of the digital circuits. Hence, the power consumption plays the main role in performance, reliability and cost of digital systems and now is a leading design constraint in microprocessor designs, especially in low-end embedded system market. It is a common trend to use ASIP in embedded systems. It provides much more flexible solutions than approaches based-on Application-specific integrated circuits (ASICs) and it is also much more efficient than standard processors in terms of both performance and power consumption [1].

The ASIP design methodology starts with extracting the data flow graph (DFG) of the application. In the next phase (i.e., identification phase) all the subgraphs that meet the constraints of the parallel hardware are enumerated as Custom Instructions (CIs). For example, the I/O ports number and propagation delays of the subgraphs are two common constraints in the CI enumeration. In the final phase, i.e. selection phase, the best CIs based-on their merit value will be selected between candidates. The merit value of each CI shows the quantity of the parameters that desired ASIP intends to improve [2]. In ASIPs, adding custom instructions to the baseline processor and customization of data path leads to decrease in the number of accesses to the instruction cache. This results in average cache access time reduction. On the other hand, increasing the cache size is a solution to increase the performance of processor. Although this improves the speed of the processor, but deteriorates the power consumption. Due to the increase in the cache size, they consume a huge deal of processor power. For instance, in Intel's StrongARM processor, caches consume more than 40% of total chip power with 27% being devoted to the instruction cache [3]. In addition, continuing advances in semiconductor technology will lead to increase transistor count in the on-chip caches, and the fraction of the total chip energy consumed by caches is likely to go up [6]. Smaller caches have shorter access time but they suffer more from miss penalty. Although this is application-dependent, finding an optimum cache size not only can improve performance, but also can reduce power consumption.

In ASIPs, the CIs reduce the number of accesses to the cache which improves the performance of the processor. Additionally, this reduction has an impact on the power consumption, and also the energy usage of the ASIP. Until now, the main focus of researchers has been on improving the speed gain of the ASIP, whereas the problem of power consumption is getting more critical in nanometer design. Therefore, the power consumption and energy usage of the ASIP designs must be considered in the design methodology. In this paper, we do a complete study on the energy usage of the cache in an extensible processor, and also, investigate how the CIs affect the access time and the energy consumption of the cache. Furthermore, we show that the extensible processor needs less cache capacity compared to the baseline processor. This motivates us to use smaller instruction cache to save energy. In section II, related works on ASIP design methodology are discussed. Section 3 demonstrates how ISA extension affects cache parameters and section IV discusses the experimental results. Finally we conclude our work in section V.

## II. RELATED WORK

In what follows, we introduce some related work on ASIP design automation. The authors in [4] have analyzed the effects of different read and write ports of register file on performance improvement and also energy consumption in extensible processors. Pozzi et al. [9] did a research on the power consumption of extensible processors. They had proposed a micro-architectural power estimation tool for customizable processors in order to monitor the power benefits of instruction set extension. Unlike the prior techniques which estimate the energy consumption of an ASIP, their methodology automates the power model construction for the customizable instructions.

Most of the studies in the field of ASIPs have focused on improving the performance of the processor among which improving CI identification [5] and selection phase [6][7][8] have attracted more attention. In [5]an enumeration algorithm was presented which reduces maximal subgraph enumeration to clique enumeration problem. Sun et al. [7] developed a branch and bound algorithm for the problem of CI selection. Cong et al. [8] considered the NP-complete CI selection as a 0-1 Knapsack problem while in [10] this problem has been formulated as an Integer Linear Programming problem. In [11] a technique for multi-cycle pipelined instruction set extension (ISE) execution was developed to solve the limitation in number of the read/write ports of the register files; under this execution model, there is no need to limit the number of inputs and outputs of each custom instruction. Additionally, some researchers have worked on resource sharing to reduce the area requirements of ISEs by sharing common hardware resources in a shared datapath [12][13][14]. As discussed, the power consumption and energy usage of the ASIP design must be considered in the design methodology. Since caches consume a huge deal of a microprocessor system's power[3], decreasing the cache energy consumption of ASIPs, can significantly reduce overall system energy. However, to the best of our knowledge, there has been no previous published work addressing this issue.

### III. IMPACT OF THE EXTENDED ISA ON CACHE PARAMETERS

In this section, we introduce our proposed method. We start by examining the impacts of CIs on number of cache accesses and then on the energy consumption. Finally we tend to reduce the cache size in order to minimize the overall energy consumption while meeting performance constraint. Each step of the proposed method is described in more details.

#### A. Impact of ISA extension on cache access numbers

First we investigate the impact of adding CIs on cache memory. As discussed, extending processor ISA reduces the number of cache memory accesses. This originates from the fact that each CI has been composed of *N* instructions; by using a CI, *N* instructions are replaced by a single custom instruction, which significantly decreases the number of required instructions (i.e. *N-1* instructions) leading to a decrease in number of cache accesses. This reduction in the number of cache accesses is mostly due to decreases in the number of cache hits. However, the number of cache misses has no significant reduction. This phenomenon is related to the fact that if there is no CI, when CPU calls an instruction which is not in cache, a cache write happens and the missing instruction along with its following instructions will be brought to the cache. Therefore there will be a cache miss and *B-1* cache hits, where *B* indicates the cache block size. When *N* instructions are replaced by only one instruction, there is only one access to cache and no more successive hits.

This means that by using CIs, the number of cache hits decreases while the number of cache misses does not change significantly. Figure 1 depicts this behavior.

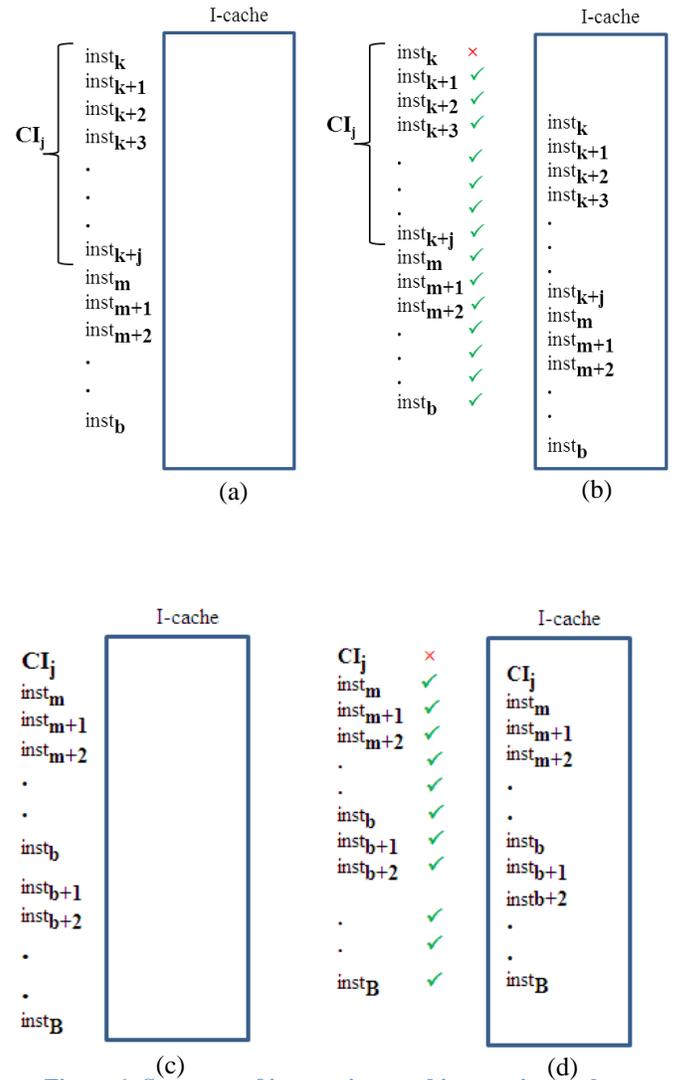

Figure 1. Sequence of instructions and instruction cache

Suppose that $CI_j$ is comprised of *j* instructions, from instruction $inst_k$ to $inst_{k+j}$ and in the best case, $inst_k$ address is exactly the same as the beginning address of the cache block. Figure 1.a shows the time when CPU calls $inst_k$ which is not in the instruction cache. It leads to a cache miss which brings a block of instruction to the cache (shown in Figure 1.b). It can be realized that in this case, we have one cache miss (because of $inst_k$) but *B-1* cache hits. We have assumed that CPU calls all *B* instructions successively and the size of our CI does not exceed the block size. Moreover, suppose that we have substituted instructions $inst_k$ to $inst_{k+j}$ with $CI_j$ and in the best case, $inst_k$ address is in the beginning. Figure 1.c and Figure 1.d depict the same phenomena while the only difference that $CI_j$ has been inserted instead of the first *j* instructions. In this case, $CI_j$ is missed. In Figure 1.d $CI_j$ along with its following instructions are brought to the cache. As shown in Figure 1.d, in this case again we have 1 cache miss but *B-j* cache hits. In this example, adding $CI_j$ does not change the number of cache misses but decreases the number of cache hits by *j-1*. Note that, if inst$_k$ address is not the same as the first address of the cache block, or in the worst case, its address is equal to the last address of the block, number of reduced cache hits will be less than *j-1*. Hence, in general, the amount of the hits and misses depends on the placement of the inst$_k$ to inst$_{k+j}$ inside the block.

Furthermore, most of the times, the CIs belong to the hot spot parts which are usually called more than other parts of the application. Hence, after they are fetched to the cache, they will be used many times. In the first access they cause a cache miss while the next accesses, will be cache hits. Due to the fact that CIs are most of the times selected form the hotspot parts of the application, adding CIs has a minor effect on the number of cache misses compared to cache hits.

### B. Impact of adding custom instructions on cache energy consumption.

Since a significant portion of processor energy consumption belongs to the cache [15], energy efficient cache architecture is a critical issue in the design of microprocessors for embedded systems. Cache energy consumption is directly proportional to the number of accesses to the cache; therefore, diminishing number of cache accesses is translated to processor energy consumption reduction. For this purpose, first we calculate the energy which is only consumed as a result of cache accesses under different cache sizes. We use equation (1) to calculate this energy consumption.

$$Energy = total\_accesses \times Energy_{port-access} \quad (1)$$

In equation (1) the *total_accesses* shows total instruction cache accesses. Also, the $Energy_{port\ access}$ is calculated by equation (2). In this equation the *hits* and *misses* are the number of the cache hit and cache miss, respectively. Also, the $hit_{Energy}(miss_{Energy})$ is the energy that is needed to fetch the instruction when the instruction is hit (missed) in the cache.

In equation (1) the only parameter which varies in different cache sizes, is energy consumed per cache read and the difference between the energy consumed per cache access in various cache sizes is negligible in comparison to the number of cache accesses in different cache sizes.

$$Energy_{port-access} = hits \times hit_{Energy} + misses \times miss_{Energy} \quad (2)$$

In the case of cache miss, the missed instruction must be fetched from the RAM. Hence, in order to achieve more accuracy, we take the energy consumed per RAM access ($Energy_{RAM}$), the energy consumed in each stall ($Energy_{stall}$) and the energy required to fill the blocks ($Energy_{block\ filling}$) in the cache miss case. Hence, cache miss energy consumption can be calculated by equation (3).

$$Energy_{miss} = Energy_{RAM} + Energy_{stall} + Energy_{block\ filling} \quad (3)$$

Finally, the total energy consumption by the cache is formulated by equation (4).

$$Energy = hits * hit_{Energy} + misses \times \quad (4)$$
$$(Energy_{RAM} + Energy_{stall} + Energy_{block\ filling})$$

### C. Reducing cache size to reduce power consumption

For each application, an optimum cash size exists. Using cache size larger than this size has no effect in miss rate reduction. Using CIs means reducing the number of cache access and also, shrinking the program code size which results in optimum cache size reduction. This allows us to reduce the cache size which is not only effective in cache static power saving (by reducing the number of transistors), but also helps dynamic power reduction ([16], according to its shorter accesses). Although it should be noted that we can reduce the cache size while the required memory average access time is not violated, because even though cache access time reduces when decreasing cache size, but miss rate increases which results in performance reduction.

Average memory access time (AMAT) is formulated in equation (5). Based on the observations described above, we can only reduce the cache size when the inequality (6) holds true.

$$AMAT = hit\ rate \times hit\ latency + miss\ rate \times \quad (5)$$
$$miss\ penalty$$

$$H_{with-CI} \times hT_{with-CI} + M_{with-CI} \times mP_{with-CI} \leq$$
$$H_{no-CI} \times hT_{no-CI} + M_{no-CI} \times mP_{no-CI} \quad (6)$$

Where $H_{no-CI}$, $M_{no-CI}$, $hT_{no-CI}$ and $mP_{no-CI}$ represent hit rate, miss rate, hit latency and miss penalty when there is no CI, respectively.

$H_{with-CI}$, $M_{with-CI}$, $hT_{with-CI}$ and $mP_{with-CI}$ denote mentioned variables in ASIP implementation. Therefore if this inequality holds true, we are sure that AMAT of the smaller cache is less or equal the AMAT of the larger cache and this guarantees that we have not violated the required performance.

## IV. EXPERIMENTAL RESULTS

Simplescalar [17]was used as our simulator framework. As for the baseline processor, we assumed a 1-issue in-order processor supporting MIPS instruction set. Since often an ASIP is a small single-issue processor, with a simple architecture [18], we conservatively use sim-fast simulator for our experimentation. We modify the code of sim-fast simulator to extract our necessary statistics. We used applications from mibench[25], mediabench[24] and SNR-RT benchmark suits [26]. The rawcaudio (rawC) and rawdaudio (rawD) were selected from mediabench, lms from SNR-RT, and G721encoder (g721Enc), G721decoder (g721Dec) and sha from mibench suit.

We have used an exact algorithm for custom instruction identification and a greedy one for custom instruction selection[19]. We first set up experiments to study the impact of adding CIs on number of cache accesses. Figure 2 shows reduction in cache access numbers in 45nm technology node.

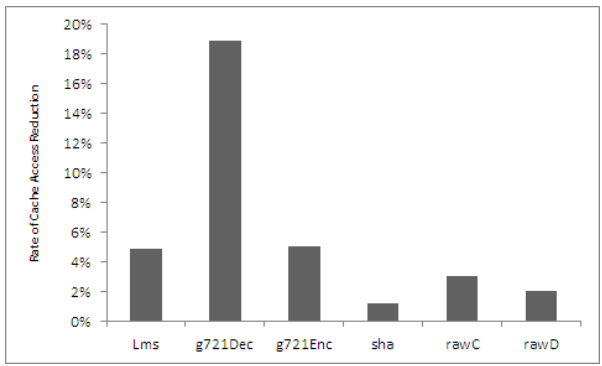

**Figure 2. Cache Access Reduction after ISA Extension**

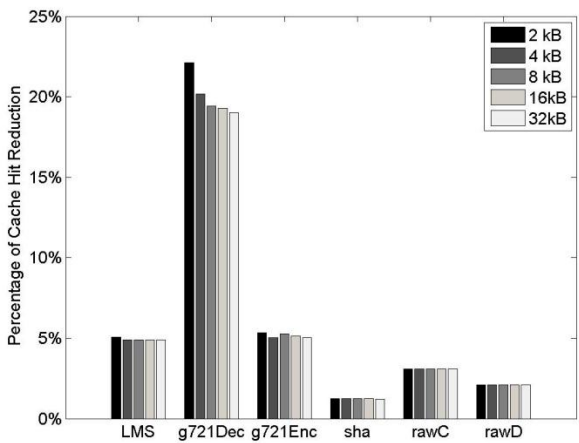

**Figure 3. Percentage of Reduction in the Number of Cache Hits**

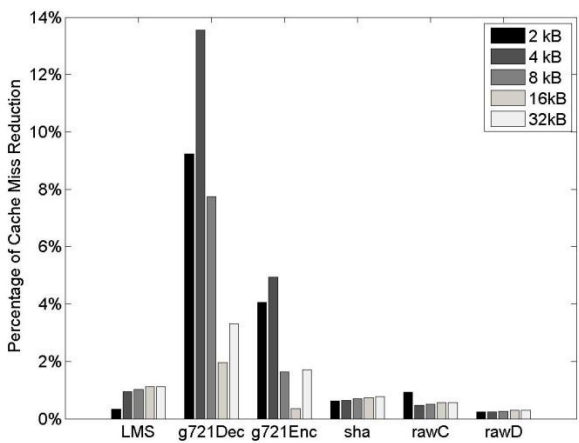

**Figure 4. Percentage of Reduction in the Number of Cache Misses**

Experimental results show that the highest cache access reduction belongs to *g721Dec* (19%) and the lowest one belongs to sha (1.2%). This difference is application dependent, i.e. program cache behavior, number of accesses to the cache memory and also complexity of CIs. Figure 3 and Figure 4 depict percentage of cache hit reduction and cache miss reduction respectively. These Figures empirically show that almost in all applications, cache hit reduction is several times higher than cache miss reduction.

To analyze the effect of adding custom instructions on cache energy consumption, we used the CACTI cache model [20] to evaluate the power consumption of each cache access. In order to calculate the values of $Energy_{stall}$, $Energy_{RAM}$ and $Energy_{block\ filling}$. [22]used the equation (4) and has shown that energy consumed per cache miss ranges from 50 to 200 times bigger than the energy consumed per cache hit [22]. We name this factor as *k_factor*. Let us assume the value of 100 for this parameter here. The simulation parameters are given in Table I. Hit latency and hit energy are produced by cacti while miss penalty and miss energy consumption are calculated by previously mentioned *k_factor*. We have used $100 * hit\_delay$ time per each miss penalty and $100 * hit\_energy$ for each miss energy consumption. Figure 5 shows the percentage of energy saving with various cache sizes. In the best, in g721Dec application when cache size is 32KB, we have reduced energy by 13.8% while the worst case relates to sha with 2KB cache size, in which energy saving is only 0.67%. In average, energy consumption has reduced by 3.7%.

**Table I. Simulation Parameter**

|  | Hit Energy (nJ) | Miss Energy (nJ) | Hit Delay (ns) | Miss Penalty (ns) |
|---|---|---|---|---|
| 1 KB | 0.00516 | 0.516 | 0.295112 | 29.5112 |
| 2 KB | 0.005368 | 0.5368 | 0.295543 | 29.5543 |
| 4 KB | 0.008101 | 0.8101 | 0.33874 | 33.874 |
| 8 KB | 0.008965 | 0.8965 | 0.347022 | 34.7022 |
| 16 KB | 0.012822 | 1.2822 | 0.366523 | 36.6523 |
| 32 KB | 0.019736 | 1.9736 | 0.406605 | 40.6605 |

One important thing to mention is that in CI identification phase, we assumed that the architecture model is a classical single issue pipelined processor which has a two-read-port and one-write-port register file, and restricted our CIs to have only two input operands which limit number and complexity of generated CIs. Since generally when the input number constraint of the custom instruction is relaxed, more operations can be clustered in one CI to exploit more parallelism that results to generate more efficient CIs [21]. Based on the observation above, if we relax this constraint in CI identification, we can generate far larger numbers of CIs and consequently get more energy saving while maintaining the original performance.

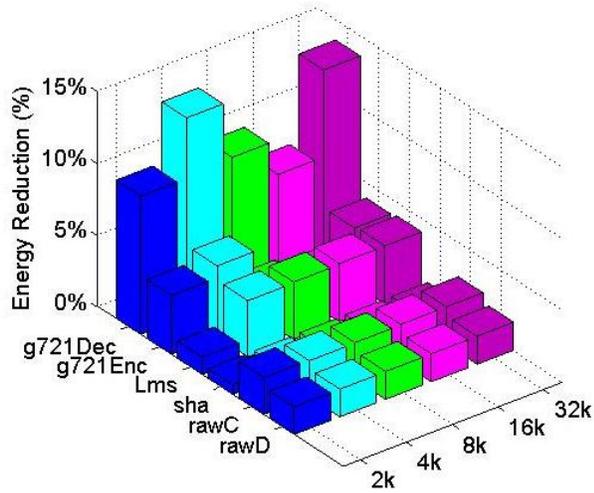

**Figure 5. Energy saving with CI extension**

**Table II. Average Memory Access Time (MILI SECOND)**

| Cache size | 1kB | 2kB | 4kB | 8kB | 16kB | 32kB |
|---|---|---|---|---|---|---|
| **Lms** | 1.986 | 0.9053 | 0.26 | 0.262 | 0.272 | 0.301 |
| **Lms2** | *1.862* | *0.8945* | *0.250* | *0.251* | *0.261* | *0.289* |
| **g721Dec** | 4.007 | 3.150 | 2.769 | 0.753 | 0.468 | 0.257 |
| **g721Dec2** | *3.382* | *2.847* | *2.386* | *0.6788* | *0.433* | *0.223* |
| **g721Enc** | 4.33827 | 3.4444 | 3.1525 | 1.296 | 0.5764 | 0.2778 |
| **g721Enc2** | *4.096* | *3.304* | *2.997* | *1.27* | *0.567* | *0.267* |
| **sha** | 0.0705 | 0.0677 | 0.0758 | 0.07203 | 0.072 | 0.0768 |
| **sha2** | *0.07* | *0.067* | *0.075* | *0.0715* | *0.071* | *0.0762* |
| **rawC** | 3.0188 | 0.238 | 0.2643 | 0.2662 | 0.2764 | 0.3056 |
| **rawC2** | *2.4594* | *0.232* | *0.258* | *0.2595* | *0.2693* | *0.298* |
| **rawD** | 0.5916 | 0.5266 | 0.601 | 0.6112 | 0.6408 | 0.7098 |
| **rawD2** | *0.5797* | *0.517* | *0.59* | *0.599* | *0.628* | *0.696* |

A larger sized cache consumes more power (both dynamic and static) than a smaller one. If an application does not need all the cache capacity, then reducing its size will save power. On the other hand, if a smaller cache results in a significant increase in the miss rate, then the savings from the smaller cache will be outweighed by the extra off-chip memory access power dissipation. Therefore while the inequality (6) holds true, we can reduce cache size to save more energy. To this end, first we calculated AMAT for different cache sizes; Table II shows AMAT of different cache sizes for a range of commercial embedded benchmarks. In the table, for each benchmark, the first row pertains to the case when no CI has been added and the second row represents this time in ASIP implementation. Our research results demonstrate that depending to the application, some cases satisfy inequality (6) and the others violate this time constraint. For example, in *g721Dec* application, we cannot switch to smaller caches but in *rawC* application, we can reduce cache size from 32 kB to 16 kB, 8kB, 4kB and 2 kB without performance loss. Such switching can be done for the other applications and for various cache sizes in a similar way. Table III shows how much dynamic energy is reduced when switching to smaller caches. In this table, $1^{st}$ size is the default size of the cache and $2^{nd}$ size is the smaller cache size to which we want to switch. For example, in sha application, when we change the cache size from 32 KB to 16 KB and 8 KB, we save 33% and 50.5% dynamic energy or in Lms application, changing the cache size from 32 to 16 and 8, results in 37.6 and 55.5% dynamic energy saving. In average, we save 28.7%, 42.6%, 53.8%, 66.7% and 69% dynamic energy when we reduce the cache size to it's $\frac{1}{2},\frac{1}{4},\frac{1}{8},\frac{1}{16}$ and $\frac{1}{32}$ size respectively.

**Table III. Dynamic Energy Reduction**

| Benchmark | $1^{st}$ Size | $2^{nd}$ Size | Energy Reduction (%) | Benchmark | $1^{st}$ Size | $2^{nd}$ Size | Energy Reduction (%) |
|---|---|---|---|---|---|---|---|
| sha | 32 | 16 | **33** | Lms | 32 | 16 | **37.6** |
| | | 8 | **50.5** | | | 8 | **55.5** |
| | | 4 | **51.7** | | | 4 | **59** |
| | | 2 | **67.3** | | 16 | 8 | **31.7** |
| | | 1 | **67.2** | | | 4 | **37** |
| | 16 | 8 | **26.7** | | 8 | 4 | **11.5** |
| | | 2 | **51.5** | | | | |
| | | 1 | **51.4** | | | | |
| | 8 | 2 | **34.4** | rawD | 32 | 16 | **36.2** |
| | | 1 | **34.2** | | | 8 | **55.1** |
| | 4 | 2 | **32.6** | | | 4 | **59.1** |
| | | 1 | **32.4** | | | 2 | **72.8** |
| rawC | 32 | 16 | **36.5** | | | 1 | **70.6** |
| | | 8 | **54.8** | | 16 | 8 | **30.9** |
| | | 4 | **58.5** | | | 4 | **37.1** |
| | | 2 | **71.6** | | | 2 | **58.2** |
| | 16 | 8 | **30.7** | | | 1 | **54.8** |
| | | 4 | **36.3** | | 8 | 4 | **10.7** |
| | | 2 | **56.5** | | | 2 | **40.6** |
| | 8 | 4 | **10.4** | | | 1 | **35.9** |
| | | 2 | **38.8** | | 4 | 2 | **34.7** |
| | 4 | 2 | **33.4** | | | 1 | **29.5** |

It is worth mentioning that this cache size reduction had no effect on the cache performance because at first, we made sure that the average access time of the smaller cash is lower than the larger one. Cache size switching was done only in the cases that inequality (6) was satisfied.

## V. CONCLUSION

In this work, the impact of CIs on the energy consumption of the instruction cache in extensible processors is studied. Our studies indicate that CI insertion can reduce the number of cache accesses tremendously which decrease the cache energy consumption. It is noteworthy to mention that according to our observation, the reduction in the number of cache accesses is mainly due to the reduction in number of cache hits. In fact, our main contribution is to reduce the number of cache accesses. Decreasing the number of cache access in turn reduces the amount of energy consumed by the cache. This fact leaded us to the idea of cache size reduction. Due to the fact that CI insertion reduces code size together with cache accesses, depending on the application, we can reduce the processor instruction cache size to save more energy. Hence we proposed analytical approach to investigate the impact of the cache size reduction on the overall energy and also the performance of the extensible processor. In average, we save 28.7%, 42.6%, 53.8%, 66.7% and 69% dynamic energy when we reduce the cache size to it's $\frac{1}{2}, \frac{1}{4}, \frac{1}{8}, \frac{1}{16}$ and $\frac{1}{32}$ size respectively.